
  \magnification\magstep1
  \baselineskip = 0.5 true cm
  \parskip=0.25 true cm

  \def\sa{\vskip 0.30 true cm}
  \def\sb{\vskip 0.60 true cm}

  \pageno = 1
  \vsize = 22.5 true cm
  \hsize = 16.1 true cm

\vglue 1 true cm

\centerline {\bf Coulomb Energy Averaged over the
    $n\ell^N$-Atomic States with a Definite Spin}

\medskip
\sa
\sb
\vskip 0.5 true cm

\centerline {M. KIBLER}

\sa

\centerline {Institut de Physique Nucl\'eaire de Lyon,}

\centerline {IN2P3-CNRS et Universit\'e Claude Bernard,}

\centerline {43 Boulevard du 11 Novembre 1918,}

\centerline {F-69622 Villeurbanne Cedex, France}

\sa
\sa

\centerline {Yu.F. SMIRNOV
\footnote\dag {Permanent address: Institute of Nuclear Physics,
Moscow State University, 119899 Moscow, Russia.}}

\sa

\centerline {Instituto de F\'\i sica,}

\centerline {Universidad Nacional Autonoma de M\'exico,}

\centerline {M\'exico D~F 001,}

\centerline {M\'exico}

\baselineskip = 0.68 true cm

\sa
\sa
\sa
\sa
\sa

\centerline {\bf Abstract}

\sa

A purely group-theoretical approach
(for which the symmetric group plays a central r\^ole),
based upon the use of properties of fractional-parentage
coefficients and isoscalar factors,
is developed for
the derivation of the Coulomb energy averaged over the states,
with a definite spin, arising from an atomic configuration
$n \ell^N$.

\sb
\sa
\sa
\sa
\sa
\sa
\sa
\sa
\sa

\noindent Published in International Journal of Quantum Chemistry
          53 (1995) 495-499.


\vfill\eject

\baselineskip = 0.99 true cm

\centerline {\bf 1. Introduction}

\sa

In the framework of an alternative parametrisation
for the theory of complex spectra [1-3],
Kibler and Katriel [2]
have conjectured a formula, viz,
$$
^{2S+1}E_{\rm av}[nl^N] = { 1 \over 2 }
N(N-1) {\Sigma} +     { 1 \over 2 }
\left [
{ N \over 2 } ( {N \over 2} + 1 ) - S(S+1)
\right ]
{\Delta}
\eqno (1)
$$
giving the average Coulomb energy
$E_{\rm av}[n\ell^N]$ for the states,
with a fixed value of the spin $S$,
which arise from an atomic configuration $n \ell^N$
of $N$ equivalent electrons in a shell $n \ell$. In Eq.~(1),
the parameters ${\Sigma}$ and ${\Delta}$ are defined through
$$
^1E_{\rm av}[nl^2] = {\Sigma} + {\Delta} \quad \qquad
^3E_{\rm av}[nl^2] = {\Sigma}
\eqno (2)
$$
corresponding to the singlet ($S \equiv S_0 = 0$) and
                     triplet ($S \equiv S_0 = 1$)
states for the case $N=2$. A formula equivalent to (1) and
(2) has been derived independently by using the method
of moments [4].

It is the aim of this short paper to give a
straightforward proof of (1)
by making use of fractional-parentage
coefficients and of an unusual sum rule satisfied by isoscalar
factors for a chain of compact groups closely connected to the
permutation group. Indeed, we work out
(in Section 2) an extension of formula (1), valid in the
general case of a spin-independent two-body interaction, and
then specialise it to the case of the Coulomb
interaction. Some concluding remarks are given in Section 3.
The relevant material (Racah's lemma and
orthogonality-completeness property for Clebsch-Gordan
coefficients or isoscalar factors) concerned with the
group-theoretical approach of Section 2 is relegated in an
appendix.

\sb

\centerline {\bf 2. Average Energies}

\sa

In the Russell-Saunders coupling (or $LS$-coupling),
the state vectors of the electronic configuration $n \ell^ N$
(for atoms and ions) may be written as
$$
\Psi \equiv \vert n\ell^{N} [f] \alpha L S M_L M_S).
\eqno\hbox{(3)}
$$
Most of the symbols in (3) have their usual meaning. Let us
simply precise that $[f]$ stands for a Young pattern,
with two columns (say of lengths $\varphi_1$ and $\varphi_2$ with
$\varphi_1 \ge \varphi_2$), which characterises
the orbital part of
$\Psi$. The spin part of $\Psi$ may be described by the
Young pattern $[\varphi] \equiv [\varphi_1 \varphi_2]$,
with two rows of lengths $\varphi_1$ and
$\varphi_2$,
which turns out to be the transposed pattern
$[{\tilde f}]$ of $[f]$. The pattern $[\varphi]$ is
(unambiguously) connected
with the number of electrons $N$ and the total
spin $S$ via
$$
\varphi_1 + \varphi_2 = N \quad \qquad
\varphi_1 - \varphi_2 = 2S.
\eqno (4)
$$
Finally, the symbol $\alpha$ in (3) collectively denotes the
remaining labels that are necessary for a single-valued
enumeration of the (allowed) atomic state vectors of the
configuration $n\ell^N$.

The vectors $\Psi$ are expressed in a basis
adapted to the following chain of groups
$$
U_{4\ell+2} \supset U_{2\ell+1} (\to SO_3 \to SO_2) \otimes
                    U_2         (\to SU_2 \to  U_1).
\eqno (5)
$$
Each of the vectors (3) spans the {\it antisymmetric} irreducible
representation class (IRC) $\left\{1^N\right\}$ of the unitary group
$U_{4\ell + 2}$. The Young patterns $[f]$ and $[\varphi]$
characterise IRC's of the groups $U_{2 \ell + 1}$ and $U_2$ for
the orbital and spin parts of $\Psi$, respectively.
The quantum numbers $L$ and $S$ refer to IRC's of the subgroups
$SO_3$ and $SU_2$ of $U_{2 \ell + 1}$ and $U_2$,
respectively. Finally, the projections $M_L$ and $M_S$ indicate
in turn
IRC's of the subgroups $SO_2$ and $U_1$ of $SO_3$ and
$SU_2$, respectively.

For a given configuration $n \ell^N$,
the total number of the vectors (3) allowed by the Pauli
principle coincides with the dimension of the IRC
$\left\{1^N\right\}$ of the group
$U_{4\ell + 2}$. It is thus given by
$\dim \left\{1^N\right\} = C^N_{4\ell + 2}$ in terms of binomial
coefficients. Among these $C^N_{4\ell + 2}$ state vectors, we
select the ones having a fixed value $S$ of the spin. Obviously,
the number of vectors $\Psi$ with $S$ fixed is nothing but the product
$N_f (2S+1)$, where $N_f$ is the dimension of the IRC $[f]$ of
$U_{2 \ell + 1}$ and $2S+1$ may be visualised as the dimension
$N_{\varphi}$ of the IRC $[\varphi]$ of $U_2$.

In the present paper, we are interested in the average energy
$$
^{2S+1}E_{\rm{av}}[n\ell^N] \; = \;
{1 \over N_f} \; {1 \over 2S+1} \;
\sum_{\alpha L} \; \sum_{M_L} \; \sum_{M_S} \;
(n\ell^N [f] \alpha L S M_L M_S \vert V \vert
 n\ell^N [f] \alpha L S M_L M_S)
\eqno\hbox{(6)}
$$
where $V$ is a spin-independent two-body Hamiltonian
$$
V = \sum_{j > i=1}^N V_{ij} \quad {\rm with} \quad
                   V_{ij} \equiv V_{ij}(r_{ij}).
\eqno (7)
$$
Equation (6) gives the average of $V$ over the $N_f (2S+1)$
state vectors (3) having a fixed spin $S$. In Eq.~(7),
the sum on $i$ and
$j$ is to be extended over the $N(N-1)/2$ two-electron
interactions $V_{ij}$~; furthermore,
$V_{ij}$ depends only on the distance $r_{ij}$ between the
electrons $i$ and $j$.

Invariance of $V$ under the rotation group $SO_3$ and the spin
group $SU_2$ ensures that (6) can be rewritten as
$$
^{2S+1}E_{\rm{av}}[n\ell^N] \; = \;
{1 \over N_f} \; \sum_{\alpha L} \;
(2L+1) \; (n\ell^N \alpha L S \vert V \vert
           n\ell^N \alpha L S).
\eqno\hbox{(8)}
$$
The next step is to introduce
coefficients of fractional parentage in order to
calculate the matrix elements of $V$
in (8). This leads to [5]
$$
\eqalign{
(n\ell^N \alpha L S \vert V \vert
 n\ell^N \alpha L S) \; = \;
{1 \over 2} \; N(N-1) \;
& \sum_{\alpha' L'  S' } \;
  \sum_{        L_0 S_0} \cr
& (\ell^N \alpha LS \{\vert \ell^{N-2} \alpha' L'S' ,
  \ell^2 L_0 S_0) \cr
& (n\ell^2 L_0 S_0 \vert V_{12} \vert n\ell^2 L_0 S_0) \cr
& (\ell^{N-2} \alpha' L' S' ,
  \ell^2 L_0 S_0 \vert\} \ell^N \alpha L S). \cr
}
\eqno\hbox{(9)}
$$
Following Racah [6, 7] and
Neudatchin and Smirnov [8],
the two-particle coefficient of
fractional parentage
$( \ell^N \left \{ \vert \ell^{N-2} , \ell^2 ) =
 ( \ell^{N-2} , \ell^2 \vert \right\} \ell^N ) ^*$
in (9) can be developed as
$$
\eqalign{
(\ell^N \alpha L S \{ \vert \ell^{N-2} \alpha' L' S' , \ell^2 L_0 S_0)
\; = \; & \sqrt{{n_{f'}n_{f_0}\over n_f}} \cr
& (\ell^N [f] \alpha L \{ \vert \ell^{N-2} [f'] \alpha' L' ,
\ell^2 [f_0] L_0) \cr
& (s^N [\tilde{f}] S \{ \vert s^{N-2} [\tilde{f'}] S' ,
s^2 [\tilde{f}_0] S_0) \cr
}\eqno\hbox{(10)}
$$
where $s=1/2$. In Eq.~(10), the IRC's $[f]$, $[f']$ and
$[f_0]$ refer
to the group $U_{2 \ell + 1}$. However, the symbols $n_f$,
$n_{f'}$
and $n_{f_0}$ denote the dimensions of $[f]$, $[f']$ and
$[f_0]$ as considered as IRC's of the permutation groups
$S_{N}$, $S_{N-2}$ and $S_{2}$, respectively. From a group-theoretical
viewpoint, the fractional-parentage coefficient
$(\ell^N \alpha L S \{ \vert \ell^{N-2} \alpha' L' S' , \ell^2 L_0 S_0)$
is identical
to the isoscalar factor
$( \{ 1^{N-2} \} \alpha' L'  S'  +
   \{ 1^2     \}         L_0 S_0 \vert
   \{ 1^N     \} \alpha  L   S)$
for the chain
$U_{4\ell+2} \supset SO_3 \otimes SU_2$ [7]. Equation
(10) then
corresponds to the factorization (see Refs.~[6] and [7])
$$
\eqalign{
( \{ 1^{N-2} \} \alpha' L'  S'  +
  \{ 1^2     \}         L_0 S_0 \vert
  \{ 1^N     \} \alpha  L   S) \; = \;
& (\{ 1^{N-2}\} [f'][\tilde{f'}] +
\{1^2\}[f_0][\tilde{f_0}]\vert
\{1^N\} [f][\tilde{f}]) \cr
& ([f']\alpha' L'+ [f_0] L_0
\vert [f]\alpha L) \cr
& ([\tilde{f'}] S' + [\tilde{f_0}]
S_0\vert [\tilde{f}] S) \cr
}\eqno\hbox{(11)}
$$
in terms of isoscalar factors for the chains
$U_{4\ell+2} \supset U_{2\ell+1} \otimes U_2$,
$U_{2\ell+1} \supset SO_3$ and
$U_{2}       \supset SU_2$.
Indeed, the correspondence between (10) and (11) yields
$$
\eqalign{
(\{1^{N-2}\} [f'][\tilde{f'}] +
\{1^2\}[f_0][\tilde{f_0}]\vert
\{1^N\}[f][\tilde{f}])
 & = \sqrt{{n_{f'}n_{f_0}\over n_f}} \cr
([f']\alpha' L' +
[f_0] L_0\vert [f]\alpha L)
 & = (\ell^N [f]\alpha L\{\vert
\ell^{N-2} [f']\alpha' L' , \ell^2 [f_0] L_0) \cr
 ([\tilde{f' }] S'  +
  [\tilde{f_0}] S_0 \vert
  [\tilde{f}]   S)
 & = (s^N [\tilde{f}] S\{\vert
s^{N-2} [\tilde{f'}] S' ,
s^2 [\tilde{f_0}] S_0). \cr
}\eqno\hbox{(12)}
$$
In Eqs.~(10) and (12), it is clear that $n_{f_0} = 1$ for both
IRC's $[f_0] = [2]$ and $[f_0] = [11]$ of $S_2$. In the case
$[f_0] = [2]$ only even values $L_0 = 0, 2, \cdots, 2 \ell$ are
admissible and $S_0=0$ while for
$[f_0] = [11]$ we have $L_0 = 1, 3, \cdots, 2 \ell - 1$ and
$S_0=1$. Furthermore in view of (12), the spin part
$(s^N [\tilde{f}  ] S \{ \vert s^{N-2} [\tilde{f'}] S' ,
  s^2 [\tilde{f}_0] S_0)$
of the fractional-parentage
coefficient (10) is trivial in the atomic case since the
isoscalar factor
$ ([\tilde{f' }] S'  +
   [\tilde{f_0}] S_0 \vert
   [\tilde{f  }] S   )$
for the chain $U_2 \supset SU_2$ is
unity for the allowed values of $S'$ and $S_0$. By combining
Eqs.~(8)-(12), we obtain
$$
\eqalign{
 ^{2S+1}E_{\rm{av}} [n\ell^N] \; = \;
& {1 \over 2} \; N(N-1)\; {1 \over N_f} \;
\sum_{\alpha L} \; \sum_{\alpha' L'} \;
\sum_{S'} \; \sum_{L_0} \; \sum_{S_0} \cr
& {n_{f'} \over n_f} \; (2L+1) \;
  (n\ell^2 L_0 S_0 \vert V_{12} \vert
   n\ell^2 L_0 S_0) \;
\vert ([f'] \alpha' L' +
       [f_0] L_0 \vert [f] \alpha L) \vert^2. \cr
}\eqno\hbox{(13)}
$$
An important step is now to effectuate the
summations over $\alpha L$ and $\alpha' L'$ in (13).
This may be easily done
by using the orthogonality-completeness relation (22) (see
appendix) applied to the
(orbital) isoscalar factors for the
chain $U_{2 \ell + 1} \supset SO_3$. We thus get
$$
\eqalign{
 ^{2S+1}E_{\rm{av}} [n\ell^N] \; = \;
& {1\over 2} \; N(N-1) \;
  \sum_{S' } \;
  \sum_{L_0} \;
  \sum_{S_0} \cr
& \Delta ([f] \vert [f'] \otimes [f_0]) \;
{n_{f'} \over n_f} \; {1 \over N_{f_0}} \; (2L_0+1) \;
(n\ell^2 L_0 S_0 \vert V_{12} \vert
 n\ell^2 L_0 S_0). \cr
}\eqno\hbox{(14)}
$$
By introducing in (14) the average energy
$$
^{2S_0+1}E_{\rm{av}} [n \ell^2] \; = \;
 {1 \over N_{f_0}} \;
\sum_{L_0} \; (2L_0+1) \;
(n\ell^2 L_0 S_0 \vert V_{12} \vert
 n\ell^2 L_0 S_0)
\eqno\hbox{(15)}
$$
for the configuration $n\ell^2$, we finally arrive at
$$
\eqalign{
 ^{2S+1}E_{\rm{av}} [n\ell^N]\; =\;
{1\over 2}\; N(N-1)\;
  \sum_{S'  \; \hbox{or} \; [f' ]} \;
& \sum_{S_0 \; \hbox{or} \; [f_0]} \cr
& ^{2S_0+1} E_{\rm{av}} [n\ell^2] \;
\Delta ([f] \vert [f'] \otimes [f_0]) \; {n_{f'} \over n_f}. \cr
}\eqno\hbox{(16)}
$$
Equation (16) provides us with a closed form expression for the
$N$-electron average energy $^{2S  +1}E_{\rm av}[n\ell^N]$
as a function of the
two-electron average energy $^{2S_0+1}E_{\rm av}[n\ell^2]$ in the case of
a general spin-independent two-body interaction.

At this stage, it is convenient
to use the $({\Sigma} , {\Delta})$-parametrization defined
by (2). In this parametrization, the energy
$^{2S+1}E_{\rm av}[n\ell^N]$ is a linear
combination of the parameters ${\Sigma}$ and ${\Delta}$.
The coefficient of            ${\Sigma}$ in this
linear
combination is clearly $(1/2)N(N-1)$. Since the parameter
${\Delta}$
appears only in the singlets of $n\ell^2$, the coefficient of
${\Delta}$ in
$^{2S+1}E_{\rm av}[n\ell^N]$ is $(1/2)N(N-1)(n_{\bar f}/n_f)$,
where $n_{\bar f}$ refers to the Young pattern $[{\bar f}]$ deduced from
$[f]$ by omiting its first row. (The latter assertion follows
from the fact that $S_0=0$ for the singlet states so that $S'=S$
and thus the sum on $[f']$ in (16) reduces to the orbital Young
pattern $[{\bar f}]$ associated            to the spin    Young
pattern $[{\bar \varphi}] \equiv [\varphi_1-1 , \varphi_2-1]$
for the spin $S$ and the number of
electrons $N-2$.)
Therefore, Eq.~(16) may be written in the form
$$
^{2S+1}E_{\rm{av}} [n\ell^N] \; = \;
{1 \over 2} \; N(N-1) \;
\left( \Sigma + {n_{\bar {f}} \over n_f} \Delta \right).
\eqno\hbox{(17)}
$$
The final step is to calculate the ratio $n_{\bar f}/n_f$ of
the dimensions of the IRC's $[{\bar f}]$ (for $S_{N-2}$)
                        and $[      f ]$ (for $S_{N  }$). This
may be achieved by using the well-known formula (see for instance
Ref.~[9]) for the IRC's of the symmetric group. We thus get
$$
{n_{\bar {f}} \over n_f} \; = \;
{\varphi_2 (\varphi_1 + 1) \over N(N-1)}.
\eqno\hbox{(18)}
$$
\vfill\eject
\baselineskip = 0.88 true cm
\noindent The introduction of (4) and (18) into (17) leads to
Eq.~(1). This completes the proof of the formula
(1) for $^{2S+1}E_{\rm av}[n \ell^N]$ in the general case of a
spin-independent two-body Hamiltonian. If we are interested in
the special case where $V$ is the repulsive Coulomb interaction
(ie, $V_{ij} = e^2/r_{ij}$), it is then sufficient to replace
the parameters $\Sigma$ and $\Delta$ in (1) by the appropriate
linear combinations of the Slater-Condon-Shortley parameters
$F^{(k)}$.

\sb

\centerline {\bf 3. Closing Remarks}

\sa

The main results of this paper [namely, Eqs.~(16) and (1)]
have been derived in the framework of atomic
spectroscopy. Let us mention that the result (1) is also of
relevance in the spectroscopy of partly-filled shell ions in
condensed matter as far as $S$ may be assumed to be
a good quantum number (an assumption that is reasonable for
transition-metal ions in crystals).
The analysis leading to the results (1) and (16)
rests on the use of the (single-configuration) shell-model and
on the assumption of a unique single-particle radial
wave-function for all the terms that the atomic (or
crystal-field) configuration gives rise to.

In the physical situation where $S$ is a good quantum number
(this situation occurs in some cases in atomic spectroscopy
and in the spectroscopy of transition-metal ions in crystals),
Eq.~(1) provides us with an expression of the barycenter
of the levels for a particular spin multiplicity~; then, the
average energy for a given $S$ corresponds to an observable
and contact with experiment may be established. Another
interest of the sum rule (1), valid even when $S$ is not a good
quantum number, is to be found in the fact that (1) is useful
for the purpose of checking matrix elements when diagonalising
the matrix of $V$ by means of an electronic computer. As a
third interest of (1), it is to be mentioned that the average
energy (1) may be used in the method of spectral
distributions
(mainly developed in nuclear physics) for the reconstruction of
the global distribution of the energy levels with a fixed value
of $S$.

Let us observe that formulas which parallel Eqs.~(16) and
(1)
might be derived in the shell model of nuclear physics. In this
connection, we may think of obtaining average energies for
fixed spin $S$, or fixed isospin $T$, or fixed spin $S$ and
isospin $T$. However, a complication arises when replacing
$N$ electrons on a $n \ell$-atomic  shell by
$N$ nucleons  on a $n \ell$-nuclear shell because the isospin
degree of freedom manifests itself by the replacement of the
   {\it trivial} spin         chain
$\left(U_2\right)_{S } \supset \left(SU_2\right)_S$ by the
{\it nontrivial} spin-isospin chain
$\left(U_4\right)_{ST} \supset
 \left(U_2 \to SU_2\right)_S \otimes
 \left(U_2 \to SU_2\right)_T$
 (cf.~Ref.~[10]). In this respect, the use of class-sum
operators [11,~12] might be of central interest.
We hope to return on this extension to nuclear physics
in a forthcoming paper.

Finally, it is to be emphasized that the unusual sum
rule (22) of the appendix
is of pivotal importance in the derivation of Eqs.~(16) and (1).
This sum rule is based on Eq.~(19)
and on the Racah lemma. It is
interesting to mention that Eq.~(19) also plays a
fundamental r\^ole in the derivation of sum rules for the intensity of
two-photon transitions between Stark levels of a transition ion
in a liquid or crystal environment [13].

\sb

\centerline {\bf Acknowledgment}

\sa

The present work was began during a stay of one of the authors
(Yu.F.~S.) in Lyon-Villeurbanne. The latter author acknowledges
the {\it Institut de Physique Nucl\'eaire de Lyon} for the kind
hospitality extended to him.

\sb

\centerline {\bf Appendix: Sum Rules for Coupling Coefficients}

\sa

Let $G$ be a finite or compact group. We use $g$ to denote
an IRC of $G$ and $\mu$ to classify the rows and columns of
a (standard) unitary matrix representation $D^g$ associated to
$g$. In this notation, the Clebsch-Gordan coefficients of the
group $G$ are written
$(g_1 g_2 \mu_1 \mu_2 \vert g_1 g_2 b g \mu )$
where $b$ is an internal multiplicity
label to be used when the Kronecker product $g_1 \otimes g_2$
contains the IRC $g$ several times. The Clebsch-Gordan coefficients
of $G$ satisfy ordinary unitarity relations controled by
summations on $\mu_1 \mu_2$ or $b g \mu$ (see for example
Refs.~[14] and [15]). In addition, they satisfy the following sum
rule (referred to as an orthogonality-completeness relation [15])
$$
\eqalign{
\sum_{\mu_1} \;
  \sum_{\mu} \; \;
  (g_1 g _2 \mu_1 \mu _2 | g_1 g _2 b  g \mu)^* \;
& (g_1 g'_2 \mu_1 \mu'_2 | g_1 g'_2 b' g \mu) = \cr
& \Delta (g | g_1 \otimes g_2) \;
  \delta (g'_2, g_2)           \;
  \delta (\mu'_2, \mu_2)       \;
  \delta (b',b)                \;
  {{\dim g}\over {\dim g_2}} \cr
}
\eqno (19)
$$
where $\Delta(g \vert g_1 \otimes g_2)$
is 1 or 0 according to as $g$ is contained in
$g_1 \otimes g_2$ or not. In the special case where $G$ is the
group $SU_2$ (or $SO_3$), there is no need for the label $b$ and
we have $g \equiv j$ (or $\ell$) and $\mu \equiv m_j$
(or $m_\ell$)~; in this
case, Eq.~(19) is a simple rewriting of one of the two
ordinary unitarity relations for the Clebsch-Gordan
coefficients $(j_1 j_2 m_1 m_2 \vert j_1 j_2 j m)$ that follows by
using the symmetry property of the latter coefficients under the
interchange $j_2 \leftrightarrow j$.

Let us now consider a subgroup $H$ of $G$. The label $\mu$ may
then be replaced by the triplet $ah\gamma$, where $h$ stands
for an IRC of the group $H$, $\gamma$ for an index to
characterise the rows and columns of
a (standard) unitary matrix representation $D^h$ associated to
$h$, and $a$ for an external multiplicity label to
be used when the IRC $h$ of $H$ occurs several times in the
reduction of the IRC $g$ of $G$. In the $G \supset H$ basis,
Eq.~(19) may be transcribed as
$$
  \eqalign{
  \sum_{a_1h_1\gamma_1} \;
  \sum_{a  h  \gamma  } \; \;
  (g_1g _2 a_1h_1\gamma_1a _2h _2\gamma _2 | g_1g _2 b  g a h \gamma)^* \;
& (g_1g'_2 a_1h_1\gamma_1a'_2h'_2\gamma'_2 | g_1g'_2 b' g a h \gamma) = \cr
 \Delta (g | g_1 \otimes g_2) \;
 \delta (g'_2, g_2)           \;
&\delta (a'_2, a_2)           \;
 \delta (h'_2, h_2)           \;
 \delta (\gamma'_2, \gamma_2) \;
 \delta (b',b)                \;
 {{\dim g}\over {\dim g_2}}. \cr
}
\eqno (20)
$$
The orthogonality-completeness relation (20) can be rewritten in
terms of isoscalar factors for the chain $G \supset H$. For
this purpose, we use the Racah factorisation lemma [6]
$$
 (g_1g_2a_1h_1\gamma_1a_2h_2\gamma_2 | g_1g_2 b g a h \gamma ) =
  \sum_{\beta} \;
  (g_1 a_1 h_1 + g_2 a_2 h_2 | b g a \beta h) \;
  (h_1 h_2 \gamma_1 \gamma_2 | h_1 h_2 \beta h \gamma)
\eqno (21)
$$
that gives an expression of a Clebsch-Gordan coefficient of $G$, in a
$G \supset H$ basis, as a linear combination of Clebsch-Gordan coefficients
$(h_1 h_2 \gamma_1 \gamma_2 \vert h_1 h_2 \beta h \gamma)$ of
$H$. The coefficients of this linear combination are the
isoscalar factors
$(g_1 a_1 h_1 + g_2 a_2 h_2 | b g a \beta h)$
for the chain $G \supset H$. (In
Eq.~(21), the label $\beta$ is a multiplicity label of type
$b$.) Then, by applying twice (21) in (20) and by using
(19) for the group $H$ in the so-obtained equation, we end up
with
$$
 \eqalign{
 \sum_{a_1h_1} \;
 \sum_{ah}\; \;
 \sum_{\beta} \;
 {{\dim h}\over {\dim h_2}} \;
 (g_1 a_1 h_1 & + g_2 a_2 h_2 | b g a \beta h)^* \;
 (g_1 a_1 h_1 + g'_2 a'_2 h_2 | b' g a \beta h) = \cr
& \Delta (g | g_1 \otimes g_2 ) \;
  \delta (g'_2, g_2)            \;
  \delta (a'_2, a_2)            \;
  \delta (b'  , b  )            \;
  {{\dim g}\over {\dim g_2}}. \cr
}
\eqno (22)
$$
Equation (22) constitutes an orthogonality-completeness
relation for the isoscalar factors of the chain $G \supset H$.

\vfill\eject

\centerline {\bf Bibliography}

\sa
\sa

\baselineskip = 0.75 true cm

\item {[1]} M.R. Kibler, Int. J. Quantum Chem. {\bf 9}, 421 (1975).

\item {[2]} M. Kibler and J. Katriel, Phys. Lett. {\bf 147A}, 417 (1990).

\item {[3]} M. Kibler and A. Partensky, in {\it Symmetry and Structural
Properties of Condensed Matter}, W. Florek, T. Lulek and M.
Mucha, Eds. (World Scientific, Singapore, 1991), p.~251.

\item {[4]} J. Karwowski and M. Bancewicz,
            J. Phys. A: Math. Gen. {\bf 20}, 6309 (1987).

\item {[5]} G. Racah, Phys. Rev. {\bf 63}, 367 (1943).

\item {[6]} G. Racah, Phys. Rev. {\bf 76}, 1352 (1949).

\item {[7]} G. Racah, {\it Group Theory and Spectroscopy}
(Institute for Advanced Study, Princeton, 1951).

\item {[8]} V.G. Neudatchin and Yu.F. Smirnov, {\it Nucleon Clusters
in Light Nuclei} (Nauka, Moscow, 1968). (In Russian)

\item {[9]} R. Pauncz, {\it Spin Eigenfunctions~-~Construction and Use}
(Plenum, New York, 1979).

\item {[10]} H.A. Jahn and H. van Wieringen,
             Proc. Roy. Soc. {\bf A 209}, 502 (1951).

\item {[11]} J. Katriel, Int. J. Quantum Chem. {\bf 35}, 461 (1989);
                                         Ibid. {\bf 39}, 593 (1991).

\item {[12]} R. Pauncz and J. Katriel, Int. J. Quantum Chem.
{\bf 41}, 147 (1992); J. Katriel and R. Pauncz,
                                       Int. J. Quantum Chem.
{\bf 48}, 125 (1993).

\item {[13]} M. Kibler and M. Daoud, Lett. Math. Phys. {\bf 28}, 269 (1993).

\item {[14]} D.T. Sviridov and Yu.F. Smirnov, {\it Theory of Optical Spectra
of Transition-Metal Ions} (Sciences, Moscow, 1977). (In Russian)

\item {[15]} M.R. Kibler, in {\it Recent Advances in Group Theory and Their
Application to Spectroscopy}, J.C. Donini, Ed. (Plenum, New York, 1979), p.~1.

\bye